\documentclass[twocolumn,showpacs,preprintnumbers,amsmath,amssymb,superscriptaddress]{revtex4}

\usepackage{color}
\usepackage{amsmath}
\usepackage{amssymb}
\usepackage{graphicx}
\usepackage{adjustbox}
\usepackage{units}
\usepackage{bm}
\usepackage{hyperref}
\usepackage{float}

\def\be{\begin{equation}}
\def\ee{\end{equation}}

\begin{document}

\title{Connecting few-body inelastic decay to many-body correlations:\\ a weakly coupled impurity in a resonant Fermi gas}
\author{S\'ebastien Laurent}\email{slaurent@lkb.ens.fr}
\author{Matthieu Pierce}
\author{Marion Delehaye}
\altaffiliation{Present Address: FEMTO-ST UMR CNRS 6174 - UFC/ENSMM/UTBM, 26 chemin de l'Epitaphe, 25030 Besan\c{c}on, France}
\author{Tarik Yefsah}
\author{Fr\'ed\'eric Chevy}
\author{Christophe Salomon}
\affiliation{Laboratoire Kastler Brossel, ENS-PSL Research University, CNRS, UPMC-Sorbonne Universit\'es, Coll\`ege de France, 24 rue Lhomond, 75005 Paris, France}


\begin{abstract}
We study three-body recombination in an ultracold Bose-Fermi mixture. We first show theoretically that, for weak inter-species coupling, the loss rate is proportional to Tan's contact. Second, using a $^7$Li/$^6$Li mixture we probe the recombination rate in both the thermal and dual superfluid regimes. We find excellent agreement with our model in the BEC-BCS crossover.  At unitarity where the fermion-fermion scattering length diverges, we show  that the loss rate is proportional the $4/3$ power of the  fermionic density.
Our results demonstrate that impurity-induced losses can be used as a quantitative probe of many-body correlations.
\end{abstract}

\maketitle

Understanding strongly-correlated quantum many-body systems is one of the most daunting challenges in modern physics. Thanks to a high degree of control and tunability, quantum gases have emerged as a versatile platform for the exploration of a broad variety of  many-body phenomena~\cite{bloch2008many}, such as the crossover from Bose-Einstein condensation (BEC) to Bardeen-Cooper-Schrieffer (BCS) superfluidity~\cite{zwerger2012BCSBEC}, quantum magnetism~\cite{Inguscio2016quantum} or many-body localization~\cite{choi2016exploring}.
At ultra-low temperatures, atomic vapors are metastable systems and are plagued by three-body recombination which represents a severe limitation for the study of some dense interacting systems. A prominent example is the strongly correlated Bose gas\,\cite{makotyn2014universal, chevy2016strongly} that bears the prospect of bridging the gap between dilute  quantum gases and liquid Helium. However inelastic losses can also be turned into an advantage. For instance, they can be used to control the state of a system through Zeno effect~\cite{syassen2008strong,daley2009atomic,zhu2014suppressing}, or serve as a probe of non-trivial few-body states, as demonstrated by the observation of Efimov trimers, originally predicted in nuclear physics, but  observed for the first time in Bose gases as resonances in three-body loss spectra~\cite{kraemer2006evidence}.

In this Letter, we study inelastic losses in a mixture of spinless bosons and spin 1/2 fermions with tunable interaction. We show that when the Bose-Fermi coupling is weak, the loss rate can be related to the fermionic contact parameter, a universal quantity overarching between microscopic and macroscopic properties of a many-body system with zero-range interactions~\cite{Tan2008energetics, tan2008large, olshanii2003short}. We first check  our prediction on the strongly attractive side of the fermionic Feshbach resonance, where we recover known results on atom-dimer inelastic scattering. We then turn to the unitary limit where the fermion-fermion scattering length is infinite. We demonstrate both theoretically and experimentally --\,\,with a $^6$Li/$^7$Li Fermi-Bose mixture\,\,-- that the bosons decay at a rate proportional to $n_{\rm f}^{4/3}$, where $n_{\rm f}$ is the fermion density. More generally our work shows that the decay of an impurity immersed in a strongly correlated  many-body system is a quantitative probe of its quantum correlations.

\begin{figure}
\centerline{\includegraphics[width=\columnwidth]{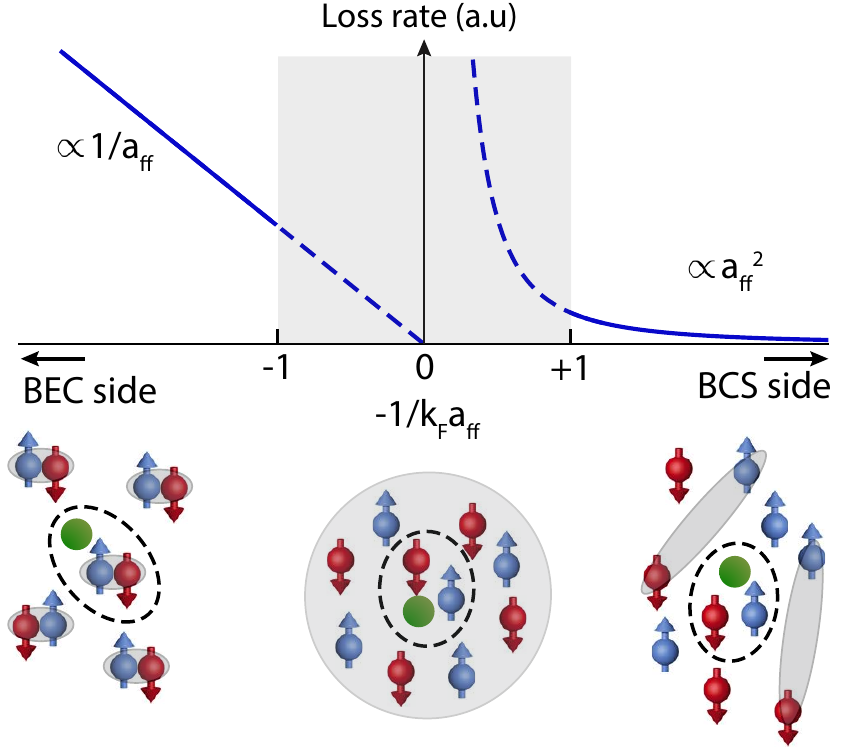}}
\caption{Sketch of inelastic decay of an impurity immersed in a tunable Fermi gas. On the BEC side, $\uparrow$ and $\downarrow$ fermions are paired in tightly bound molecules and the decay mechanism is a two-body process involving the impurity (green disk) and a molecule. The loss rate scales as $1/a_{\rm ff}$\cite{d2008suppression, zirbel2008collisional}. On the BCS side, the loss occurs through a three body-process and it scales as $a_{\rm ff}^2$ in the mean-field limit~\cite{d2008suppression}. The extrapolation of these two asymptotic behaviors towards the strongly correlated regime yields contradictory results (grey area).}
\vspace{-0.5cm}
\label{Fig0}
\end{figure}

Inelastic decay of an impurity inside a two-component Fermi gas has been studied previously both in the weakly and strongly attractive limits of the BEC-BCS crossover~\cite{ d2008suppression,dincao2009short,spiegelhalder2009collisional,Khramov2012dynamics}. When the fermion-fermion interaction is weak, the fermions behave almost as isolated particles and the recombination can be described as a three-body process involving one spin-up ($\uparrow$), one spin-down ($\downarrow$) fermion and the impurity (a boson in our experiments). In this case, the impurity/boson density $n_{\rm b}$  follows a rate equation $\dot n_{\rm b}=-L_3 n_{\rm f}^2 n_{\rm b}$, with $L_3\propto a_{\rm ff}^2$, where $a_{\rm ff}$ is the fermion-fermion scattering length~\cite{d2008suppression,zirbel2008collisional,spiegelhalder2009collisional}. In contrast, on the strongly attractive side of the Feshbach resonance, the fermions form halo-dimers of size $\simeq a_{\rm ff}$ and the relaxation occurs through two-body processes between one such molecule and one boson. In this case the rate equation for bosons reads $\dot n_{\rm b}=-L_2 n_{\rm m}n_{\rm b}$, where $n_{\rm m}=n_{\rm f}/2$ is the molecule density. Far from the Feshbach resonance the two-body loss rate scales as $1/a_{\rm ff}$ as a consequence of the enhanced overlap of the halo-dimer wavefunction with the deeply bound product molecules~\cite{d2008suppression,zirbel2008collisional}. However, these two scalings give rise to a paradox in the central region of the BEC-BCS crossover. Indeed, as depicted in Fig.\ref{Fig0}, the extrapolation towards unitarity leads to contradictory results depending on whether we approach the resonance from the BEC or the BCS side. In the former case, one would predict an increasingly long lifetime at unitarity while it tends to a vanishingly small value in the latter case. This paradox has a fundamental origin: these two scalings are obtained in the dilute limit where the recombination can be described by a well-defined few-body process,
 whereas this hypothesis  fails in the strongly correlated regime where $n_{\rm f}|a_{\rm ff}|^3\gg 1$. There, it is not possible to single out two fermions from the whole many-body system. Instead, the inelastic loss involving a boson and two fermions is tied to the correlations of the whole ensemble. A first hint towards reconciling these two behaviors near unitarity is to assume that they saturate for $a_{\rm ff}\simeq n_{\rm f}^{-1/3}$, yielding the same scaling $\dot n_{\rm b} \propto n_{\rm f}^{4/3} n_{\rm b}$.

The three asymptotic regimes -- BEC, BCS, and unitary -- were obtained using different theoretical approaches and we now show that, using Tan's contact, they can be unified within the same framework. The recombination rate is proportional to the probability of having the three particles within a distance $b$ from each other, where $b$ is the typical size of the deeply-bound molecule formed during the collision\cite{Kagan1985effect, petrov2004weakly}. Take $\rho_3(\bm r_\uparrow,\bm r_\downarrow,\bm r_{\rm b})$ the three-body probability distribution of the system. When the bosons are weakly coupled to the fermions, we can factor it as $\rho_3(\bm r_\uparrow,\bm r_\downarrow,\bm r_{\rm b})=\rho_{\rm f}(\bm r_\uparrow,\bm r_\downarrow)\rho_{\rm b}(\bm r_{\rm b})$. Integrating over the positions of the three atoms  we readily see that the three-body loss rate is proportional to Tan's Contact parameter $C_2$ of the fermions that gives the probability of having two fermions close to each-other~\cite{Tan2008energetics}.  $C_2$ is calculated using the equation of state of the system thanks to the adiabatic-sweep theorem
\be
C_2=-\frac{4\pi m_{\rm f}}{\hbar^2}\frac{\partial F}{\partial (1/a_{\rm ff})}.
\ee
where $m_{\rm f}$ is the fermion mass and $F$ is the free-energy of the fermionic gas per unit-volume~\cite{olshanii2003short,tan2008large}. The asymptotic expressions of $C_2$ in the BEC, BCS and unitary regimes are listed in Table~\ref{Table1}. In the deep BEC limit, the free energy is dominated by the binding energy of the molecules $\hbar^2/m_{\rm f}a_{\rm ff}^2$; in the BCS regime $C_2$ is derived using the mean-field approximation~\cite{Tan2008energetics}.
At unitary, the expression of the contact stems from the absence of any length scale other than the inter-particle distance. The dimensionless parameter $\zeta= 0.87(3)$ was determined both theoretically~\cite{astrakharchik2004eq} and experimentally~\cite{navon2010equation,sagi2012measurement,kuhnle2011temperature}. Expressions listed in Table \ref{Table1} confirm that the contact parameter and the bosonic loss rate follow the same scalings with density and scattering length.

\begin{table}[t!]
\begin{tabular}{c|ccc}
&\makebox[2cm]{BEC}&\makebox[2cm]{Unitary}&\makebox[2cm]{BCS}\\
\hline
\hline\\[-2ex]
$C_2$&$8 \pi \dfrac{n_{\rm m}}{a_{\rm ff}}$&$\dfrac{2\zeta}{5 \pi}k_F^4$&$4\pi^2 a_{\rm ff}^2n_{\rm f}^2$\\[2ex]
$\dfrac{\dot n_{\rm b}}{n_{\rm b}}$
&
$\propto \dfrac{n_{\rm m}}{a_{\rm ff}}$\cite{d2008suppression}
&$\propto n_{\rm f}^{4/3}$&$\propto a_{\rm ff}^2 n_{\rm f}^2$\cite{d2008suppression}\\[2ex]
\hline
\end{tabular}

\caption{Scaling of Tan's contact~\cite{Tan2008energetics} and of the boson/fermion mixture loss rate in the BEC-BCS crossover. Both scalings are identical in the weakly and strongly attractive limits. As $k_{\rm F}=(3\pi^2 n_{\rm f})^{1/3}$, at unitarity $C_2$ scales as $n_{\rm f}^{4/3}$. $\zeta$ is a dimensionless constant, $\zeta=0.87(3)$~\cite{navon2010equation,hoinka2013}.} \label{Table1}
\end{table}

We support this relationship between inelastic losses and Tan's contact by considering a microscopic model where the recombination is described by a three-body Hamiltonian
\be
\begin{split}
\widehat H_3=&\int d^3\bm r_{\rm b} d^3\bm r_{\uparrow}d^3\bm r_{\downarrow} g(\bm r_{\rm b},\bm r_\uparrow,\bm r_\downarrow)\times \\
&\widehat\Psi_{\rm m}^\dagger\left(\frac{\bm r_\uparrow+\bm r_\downarrow}{2}\right)\widehat\Psi_{\rm b}^\dagger(\bm r_b)\widehat\Psi_{\rm b}(\bm r_{\rm b})\widehat\Psi_{\uparrow}(\bm r_\uparrow)\widehat\Psi_{\downarrow}(\bm r_\downarrow)\\
&+{\rm h.c.},
\end{split}
\ee
where $\widehat\Psi_\alpha$ is the field operator for the species $\alpha$ and the coupling $g$ takes significant values only when the three particles are within a distance $b$\footnote{Note that this Hamiltonian describes the formation of fermion-fermion molecules but our conclusion remains valid in the Bose-Fermi case.}.
Assuming that  $b$ is the smallest distance scale in the problem and that this Hamiltonian can be treated within Born's approximation we find that (see~\cite{SuppMat})
\be
\dot n_{\rm b}=-\gamma C_2 n_{\rm b},
\label{Eq1}
\ee
 The constant $\gamma$ depends on the coupling $g$ and describes the coupling to deeply bound non-resonant states; hence  $\gamma$ has essentially no variation with magnetic field across the fermionic Feshbach resonance.

Eq.~(\ref{Eq1}) is the main prediction of this Letter and we explore the consequences of this equation by measuring the lifetime of an ultracold Fermi-Bose mixture of $^6$Li and $^7$Li atoms. Our experimental setup is described in~\cite{Ferrier2014Mixture}. The $^6$Li atoms are prepared in a  spin mixture $\uparrow, \downarrow$ of $|F=1/2,m_F=\pm1/2\rangle$ for which there is a broad Feshbach resonance at 832\,G~\cite{Zurn:2013}. The $^7$Li atoms are transferred into  the $|F=1,m_F=0\rangle$ featuring two Feshbach resonances, a narrow one at 845.5\,G and a broad one at 893.7\,G~\cite{SuppMat}. The scattering length between bosons and fermions is $a_{\rm bf}=40.8\,a_0$ and is equal for the $\uparrow, \downarrow$ states. It can be considered constant in the magnetic field range that we explored, 680-832\,G.  The atoms are confined in a hybrid magnetic/optical trap and are evaporated at the $^6$Li Feshbach resonance until we reach dual superfluidity or any target temperature. We ramp the magnetic field to an adjustable value in 200\, ms and wait for a variable time $t$. We then measure the atom numbers of the two species by in situ imaging or after time of flight.
\begin{figure}
\centerline{\includegraphics[width=\columnwidth]{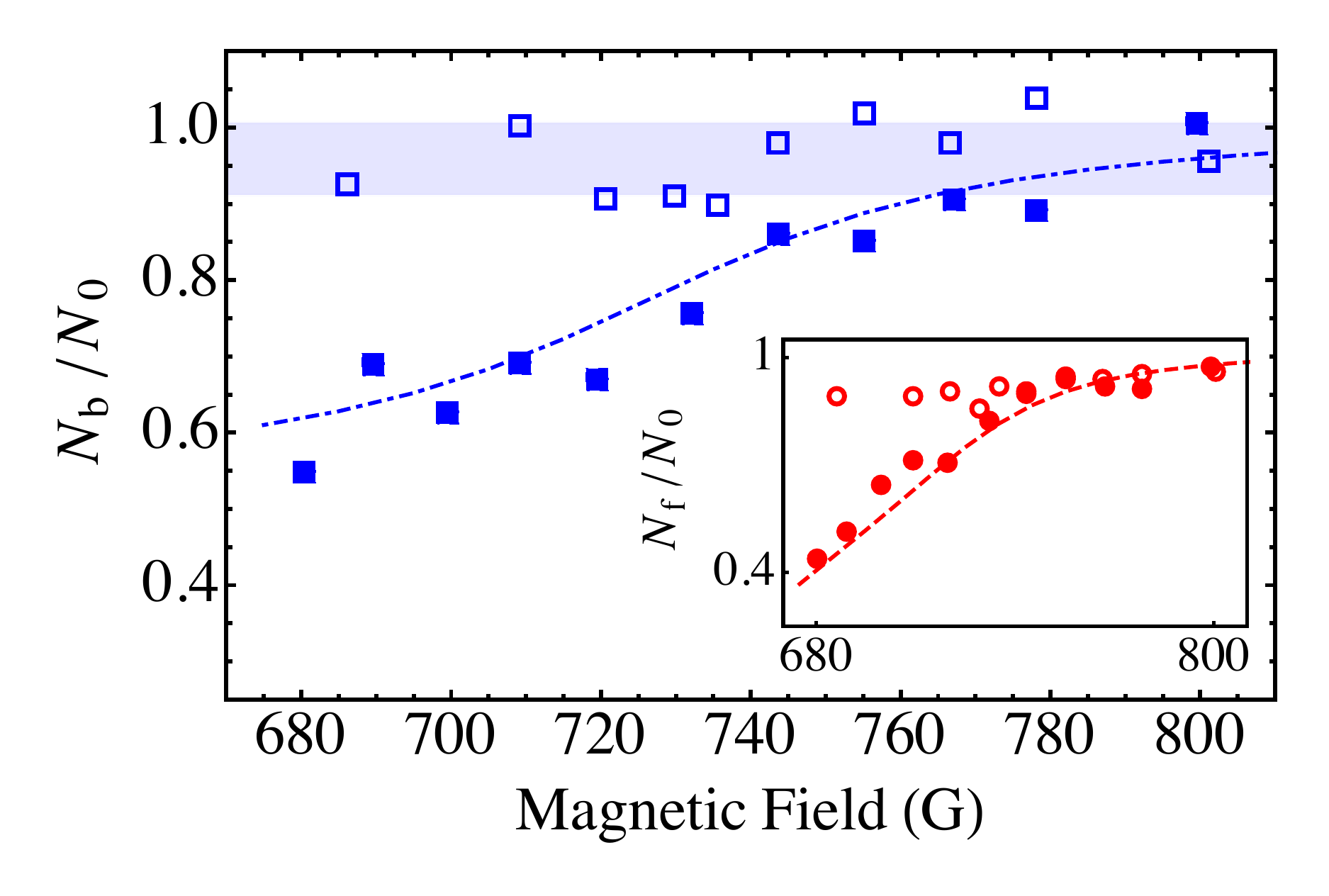}}
\caption{Remaining fraction of bosons (blue symbols) and fermions (red symbols, inset) after a 1\,s and 1.5\,s waiting time for spin-balanced (filled symbols), resp.  $90\%$ polarized (open symbols) fermions. The  blue dash-dotted (red dashed, inset) curve is a coupled loss model describing the competition between boson fermion-dimer decay ($\propto 1/a_{\rm ff}$) and dimer-dimer decay ($\propto 1/a_{\rm ff}^{2.55}$)~\cite{petrov2004weakly,SuppMat}. The blue-shaded area represents the $1\sigma$ fluctuations for the remaining fraction of bosons with spin-polarized fermions.
The initial atom numbers are $3\times 10^5$\, for $^6$Li and $1.5\times 10^5$ for $^7$Li at a temperature $T\simeq 1.6\,\mu$K with trap frequencies $\nu_z=26\,$Hz and $\nu_r=2.0\,$kHz.}
\label{Fig1}
\end{figure}

We first show that the dominant boson loss mechanism on the BEC side of the resonance involves one boson, one fermion $\uparrow$, and one fermion $\downarrow $. This is easily done by comparing the boson losses for spin-balanced and spin-polarized fermionic samples. Fig.~\ref{Fig1} displays the remaining fraction of bosons and fermions after a waiting time of 1~s for balanced fermions and 1.5~s for spin-polarized fermions with $90\,\%$ polarization. We observe that the losses for high spin polarization are strongly suppressed indicating that fermions of both spin components are required to eliminate one boson.

\begin{figure}
\centerline{\includegraphics[width=\columnwidth]{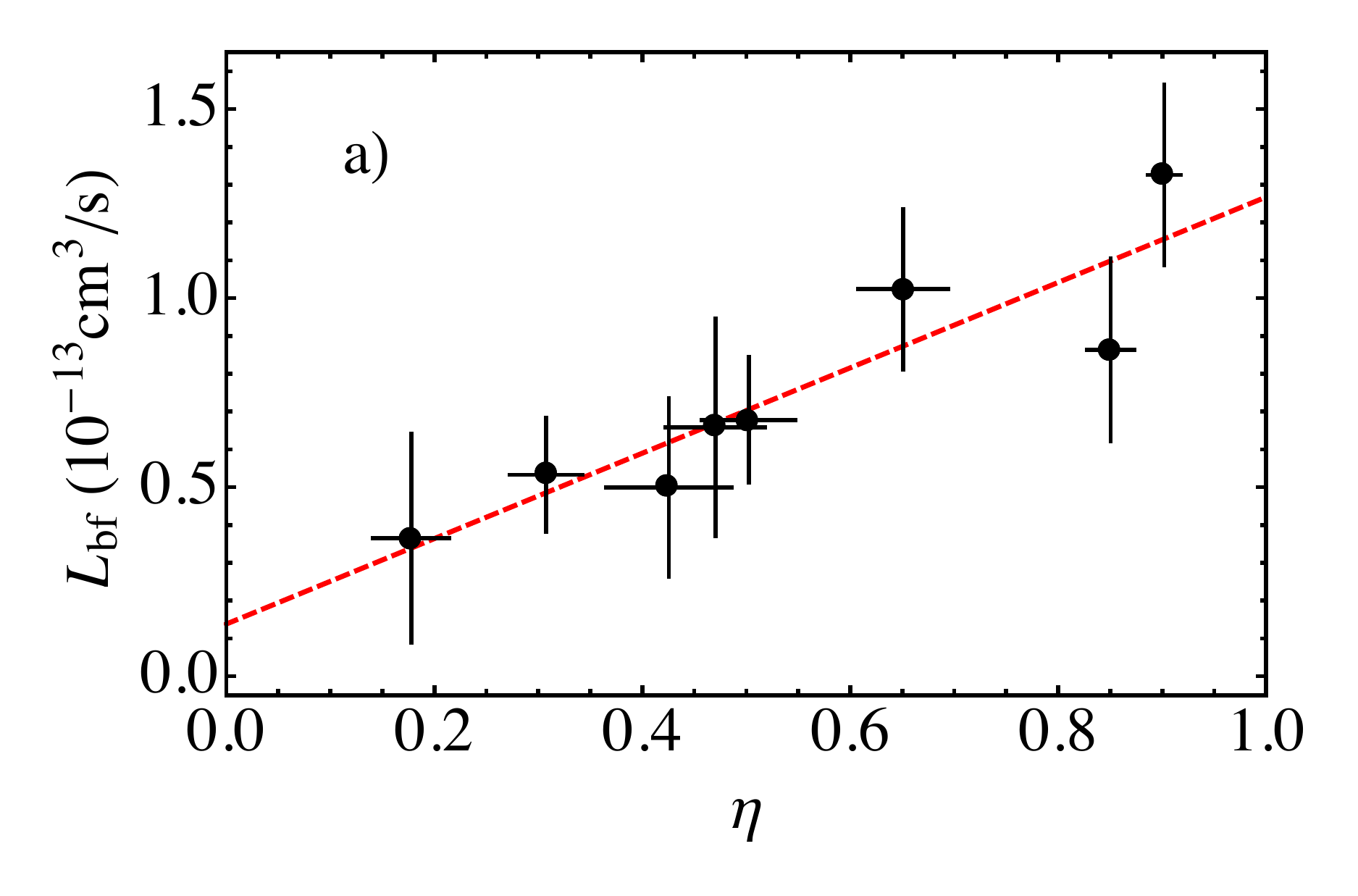}}
\centerline{\includegraphics[width=\columnwidth]{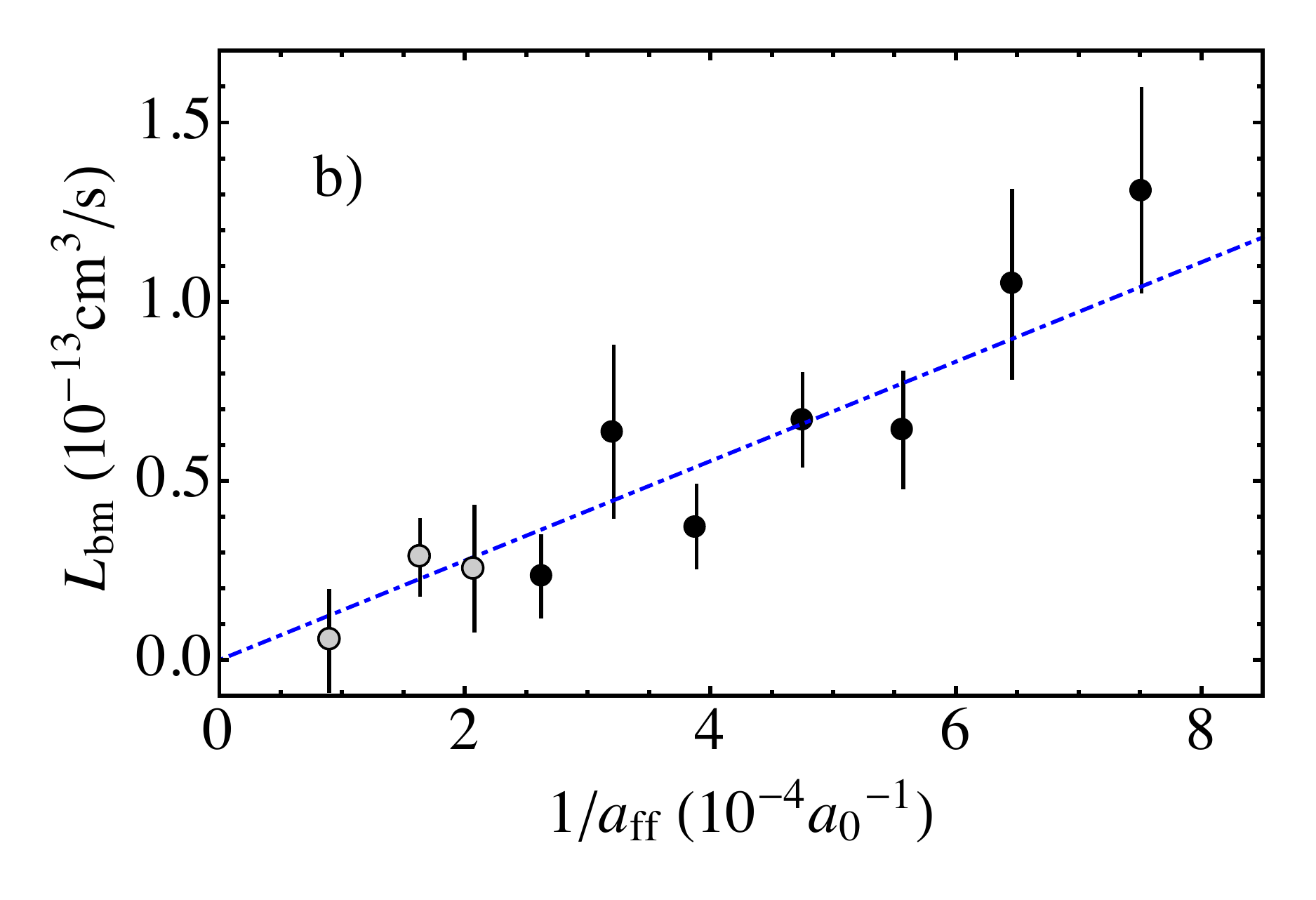}}
\caption{a) Boson-fermion loss rate vs molecule fraction. Circles: experimental data. The vertical error bars represent the statistical errors for $L_{\rm bf}$ from fitting the loss curves. The horizontal error bars represent the statistical errors on the molecule fraction due to $^6$Li number fluctuations. The red dashed line is a linear fit to the data. b)  Boson-dimer loss rate vs inverse scattering length. The blue dot-dashed line is a linear fit to the data with $n_{\rm f}a_{\rm ff}^3\leq 0.025$ (black circles),  providing $\gamma=1.17(11)\times 10^{-27}\rm\,m^{4}.s^{-1}$, see Eq. \ref{Eq1}.}
\label{Fig2}
\end{figure}

Second we show that the losses in the weakly interacting regime $na_{\rm ff}^3 \ll 1$ (deep BEC side of the resonance, 720\,G)  are proportional to the fraction of molecules in the sample, $\eta =2N_{\rm m}/(N_{\rm f}+2N_{\rm m})$. This fraction is varied by changing the temperature from $1\, \mu$K to $4\,\mu$K and $^6$Li densities from $2\times 10^{12}\,\rm{cm}^{-3}$ to $1.0\times 10^{13}\,\rm{cm}^{-3}$. In these temperature and density ranges, both gases are well described by Maxwell-Boltzmann position and velocity distributions.
The molecular fraction is calculated using the law of mass-action~\cite{Chin2004thermal, SuppMat} and is assumed to be time-independent owing to the high formation rate of halo-dimers ($\simeq \hbar a_{\rm ff}^4/m_{\rm f}$)\cite{fedichev1996three}.
We  extract the inter-species decay rate by fitting the time evolution of the bosonic population

\begin{eqnarray}
\dot N_{\rm b}&=&- L_{\rm bf}\langle n_{\rm f}\rangle N_{\rm b}-\Gamma_{\rm v} N_{\rm b}.
\end{eqnarray}
where  $\langle\cdot\rangle$ represents the trap-average, and $\Gamma_{\rm v}$ is the one-body residual gas loss rate ($0.015\,{\rm s}^{-1}$).

The data in Fig.~\ref{Fig2}a  shows that the boson loss-rate is proportional to the molecule fraction of the fermionic cloud. Introducing the boson-fermion dimer molecule loss rate $L_{\rm bm}$ defined by $L_{\rm bm}\langle n_{\rm m}\rangle=L_{\rm bf}\langle n_{\rm f}\rangle$,
 we check the proportionality of $L_{\rm bm}$ with $1/a_{\rm ff}$ predicted in table \ref{Table1}  by repeating the loss measurements for different magnetic fields in the interval 690-800\,G, see Fig.~\ref{Fig2}b. From a linear fit to the data where interaction effects are negligible ($n_{\rm f}a_{\rm ff}^3\leq 0.025$), we extract the slope  $\gamma=1.17(11)\times 10^{-27}\,\rm m^{4}.s^{-1}$ entering in Eq.~(\ref{Eq1}).

Since $\gamma$ doesn't depend on the magnetic field we can now predict the loss rate anywhere in the BEC-BCS crossover using Eq.~(\ref{Eq1}). The strongly interacting unitary regime ($1/a_{\rm ff}=0$) is particularly interesting and we measure the boson decay rate at 832\,G in the low temperature dual superfluid regime~\cite{Ferrier2014Mixture}.  The mixture is initially composed of about $40\times 10^3$ fully condensed $^7$Li bosons and $150\times 10^3$ $^6$Li spin-balanced fermions at a temperature  $T\simeq 100\, \rm nK$ which corresponds to $T/T_{\rm F}\simeq 0.1$ where $T_{\rm F}$ is the Fermi temperature.
At this magnetic field value, the atoms are now closer to the boson Feshbach resonance located at 845.5\,G and bosonic three-body losses are no longer negligible.
The time dependence of the boson number is then given by
\be
\dot N_{\rm b}=- L_{\rm b}\langle n_{\rm b}^2\rangle N_{\rm b}-\Gamma_{\rm bf} N_{\rm b}-\Gamma_{\rm v} N_{\rm b}.
\label{Eq2}
\ee

To extract $\Gamma_{\rm bf}$ we measure independently $L_{\rm b}$ with a BEC without fermions in the same trap and inject it in Eq.~(\ref{Eq2}), see~\cite{SuppMat}. We typically have  $L_{\rm b}\langle n_{\rm b}^2\rangle=0.1-0.4\, \rm s^{-1}$, and $L_{\rm b}=0.11(1)\times 10^{26}\,\rm cm^6.\,s^{-1}$ consistent with the model of~\cite{Shotan2014three}.
Repeating such measurements for different fermion numbers and trap confinement, we now test the expected $n_{\rm f}^{4/3}$ dependence of the Bose-Fermi loss rate at unitarity (central column in Table\,\ref{Table1}). In this dual superfluid regime, the size of the BEC is much smaller than that of the fermionic superfluid  and the BEC will mainly probe the central density region $n_{\rm f}(r=0)$. However,  it is not truly a point-like probe, and introducing the ratio $\rho$ of the Thomas-Fermi radii for bosons and fermions, we obtain the  finite size correction for  Eq.~(\ref{Eq1})\,\cite{SuppMat}:

\begin{figure}
\centerline{\includegraphics[width=\columnwidth]{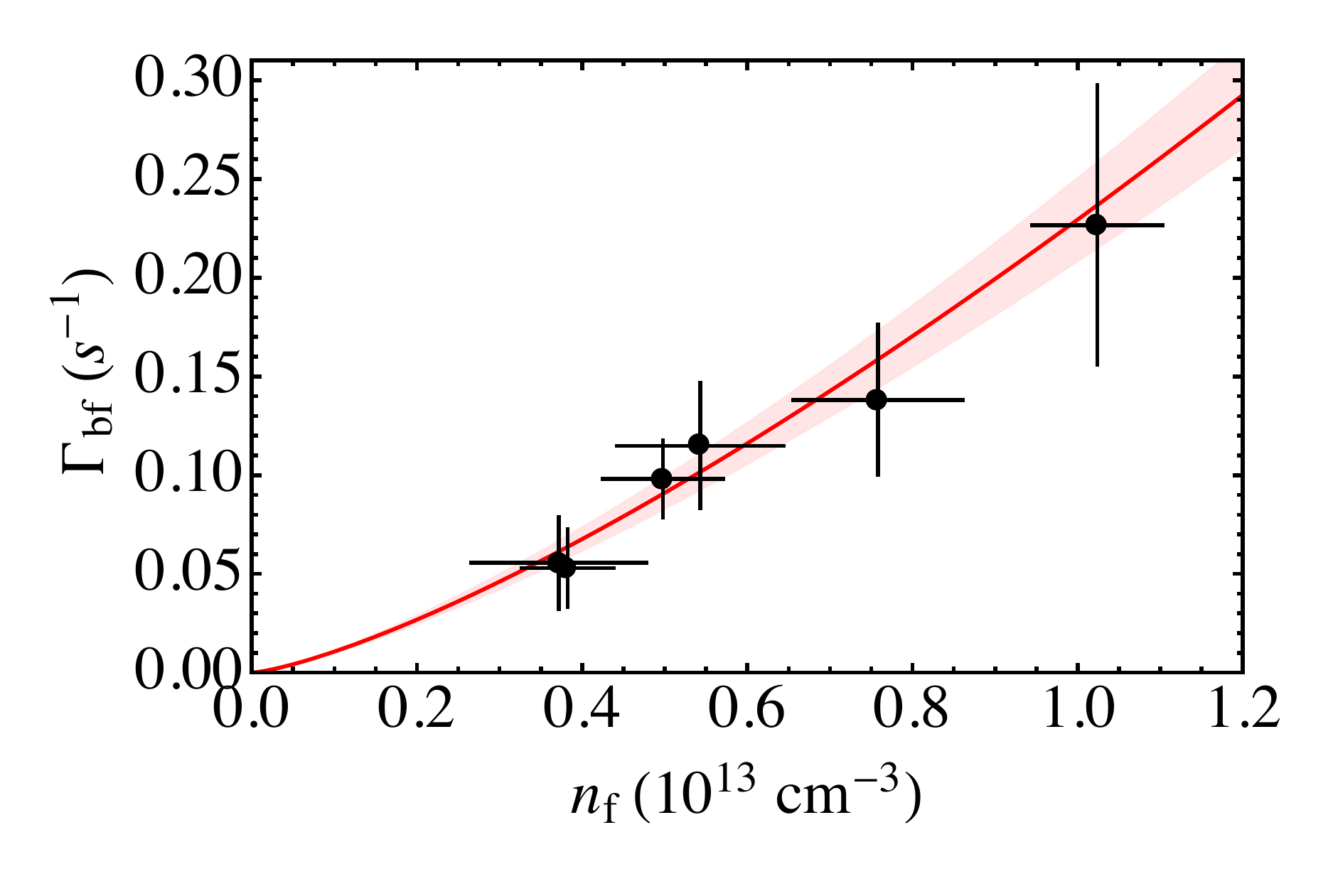}}
\caption{Boson loss rate versus fermion central density at unitarity, $n_{\rm f}=n_{\rm f}(0)$. Circles: experimental data. The red line is the $n_{\rm f}^{4/3}$ prediction of Eq.~(\ref{EquationQuiTue}) without any adjustable parameter. The red shaded area represents the 1\,$\sigma$ uncertainty resulting from the error on $\gamma$.}
\label{Fig4}
\end{figure}

\be\Gamma_{\rm bf}=\gamma\,C_2(0)\,(1-\frac{6}{7}\rho^2),\label{EquationQuiTue}\ee
where $C_2(0)=\frac{2\zeta}{5\pi\xi}(3\pi^2n_{\rm f}(0))^{4/3}$, $\xi=0.376(4)$ is the Bertsch parameter\,\cite{ku2012revealing}, and the last factor in parenthesis amounts to 0.9. The prediction of Eq.~(\ref{EquationQuiTue}) is plotted as a red line in Fig.\,\ref{Fig4} and is in excellent agreement with our measurements without any adjustable parameter.
 Alternatively, a power-law fit $A n^p$ to the data yields an exponent $p=1.36(15)$ which confirms the $n_{\rm f}^{4/3}$ predicted scaling at unitarity. Finally fixing $p$ to $4/3$ provides the coefficient $A$ and a value of the homogeneous contact $\zeta=0.82(9)$ in excellent agreement with previous measurements, $\zeta=0.87(3)$~\cite{navon2010equation, hoinka2013}. This demonstrates that impurity losses act as a microscopic probe of many-body correlations.

The bosonic or fermionic nature of the probe is of no importance. Provided the coupling between the impurity and the resonant gas is weak, our method can also be applied to other mixtures. It gives a framework to interpret the experimental data on $^6$Li/$^{40}$K~\cite{spiegelhalder2009collisional} and, in particular, to test our prediction on the BCS side of the Feshbach resonance. It can also be applied to the recently observed $^6$Li/$^{174}$Yb~\cite{roy2016two}, $^6$Li/$^{41}$K~\cite{yao2016observation} and $^6$Li/$^7$Li~\cite{Ikemachi2016all-optical} dual-superfluid Bose-Fermi mixtures. Our observation of a loss rate scaling $\propto n_{\rm f}^{4/3}$ at unitarity is in stark contrast with the generic case $n^p$, where the integer $p$ is the number of particles involved in the recombination process. A fractional exponent is also predicted to occur for the resonant Bose gas~\cite{makotyn2014universal, chevy2016strongly}.

A first extension of this work is to investigate regimes where $a_{\rm bf}\simeq a_{\rm ff} \gg n^{-1/3}$ and Born approximation breaks down. In this case Efimovian features are expected to occur~\cite{Braaten2009three, Ottenstein2008collisional}. Second, our method provides a unique microscopic way to measure the contact quasi-locally in a harmonic trap. An important perspective is to determine the homogeneous contact of the unitary Fermi gas at finite temperature, whose behavior is largely debated near the normal-superfluid transition~\cite{Sagi2012}. Our approach can also be extended to the measurement of the three-body contact which is unknown for the two component Fermi gas at unitarity~\cite{Fletcher2016,zirbel2008collisional,Du2009inelastic}.

\begin{acknowledgements}
The authors thank  G. Shlyapnikov, B. Svistunov and F. Werner for helpful discussions. They acknowledge support from R\'egion Ile de France (DIM IFRAF/NanoK), ANR (Grant SpiFBox)  and European Union (ERC Grant ThermoDynaMix).
\end{acknowledgements}

\clearpage

\newcommand{\7}{$^7$Li}
\newcommand{\6}{$^6$Li}
\newcommand{\ket}[1]{\left| #1 \right>} % for Dirac bras
\newcommand{\bra}[1]{\left< #1 \right|} % for Dirac kets
\newcommand{\bea}{\begin{eqnarray}}
\newcommand{\eea}{\end{eqnarray}}
\graphicspath{{../}}
\renewcommand{\abstractname}{}
\renewcommand{\thefigure}{S\arabic{figure}}
\renewcommand{\theequation}{S\arabic{equation}}

\centerline{\large{\bf SUPPLEMENTARY INFORMATION}}

\section{Microscopic three-body loss model}

\medskip

We derive here the relationship between the contact and the bosonic loss rate using a microscopic model. We consider the three-body Hamiltonian

\be
\begin{split}
\widehat H_3=&\int d^3\bm r_1 d^3\bm r_2 d^3\bm r_3 g(\bm\rho_1,\bm \rho_2)\times \\
 &\widehat\Psi_M^\dagger\left(\frac{\bm r_1+\bm r_2}{2}\right)\widehat\Psi_3^\dagger(\bm r_3)\widehat\Psi_3(\bm r_3)\widehat\Psi_2(\bm r_2)\widehat\Psi_1(\bm r_1)\\
 &+{\rm hc},
\end{split}
\ee
where $\widehat\Psi_\alpha$ are the field operators for the atoms ($\alpha=1,2,3$) and the molecule ($\alpha=M$). $g(\bm\rho_1,\bm\rho_2)$ is a kernel describing the molecule formation and is expressed in term of Jacobi's coordinates $\bm\rho_1=\bm r_1-\bm r_2$ and $\bm\rho_2=\bm r_3-(\bm r_1+\bm r_2)/2$. Its characteristic width is of the order of the typical size $b$ of deeply bound molecules ($\simeq$ Van Der Waals length) and is assumed to be much smaller than the other relevant length scales of the problem (scattering length and inter-particle distance).  Finally we assume for the sake of simplicity that all atomic species ($\alpha=1,2,3$) have the same mass.

We now recast $\widehat H_3$ in momentum space by taking
\be
\widehat\Psi_\alpha (\bm r)=\frac{1}{\sqrt{\Omega}}\sum_{\bm k}e^{i\bm k\cdot\bm r}\widehat a_{\alpha}(\bm k)
\ee
where $\Omega$ is a quantization volume.
We then have
\be
\begin{split}
\widehat H_3=&\frac{1}{\Omega^{3/2}}\sum_{\substack{\bm k_1,\bm k_2\\ \bm k_3,\bm k'_3}}   \tilde g((\bm k_1-\bm k_2)/2,(\bm k'_3-\bm k_3))\times\\
&\widehat a_M(\bm k_M)^\dagger \widehat a_3(\bm k'_3)^\dagger \widehat a_3 (\bm k_3) \widehat a_2 (\bm k_2) \widehat a_1 (\bm k_1) +{\rm h.c.},
\end{split}
\ee
with
$$
\tilde g (\bm q,\bm q')=\int d^3\bm\rho_1 d^3\bm\rho_2 e^{-i(\bm q\cdot\bm\rho_1+\bm q'\cdot\bm\rho_2)} g(\bm\rho_1,\bm\rho_2),
$$
and $\bm k_M=\bm k_1+\bm k_2+\bm k_3-\bm k'_3$ owing to momentum conservation.

We treat the 1-2 atoms as a strongly correlated many-body system and we neglect their interactions with the 3-atoms or the molecules. We consider the initial state $|i\rangle=|0\rangle_{12}\otimes |\bm k_3\rangle_3\otimes |\emptyset\rangle_M$ where $|0\rangle_{12}$ is the ground-state of the many-body 1-2 system, $|\bm k_3\rangle_3$ corresponds to a single 3-particle with momentum $\bm k_3$ and $|\emptyset\rangle_M$ is the molecule-vacuum. Assuming that the Hamiltonian $\widehat H_3$ can be treated perturbatively, the molecule formation-rate $\Gamma$ is given by Fermi's Golden Rule

\be
\Gamma=\frac{2\pi}{\hbar}\sum_f |\langle f|\widehat H_3|i\rangle|^2\delta (E_f-E_i).
\ee
Here, the final state takes the form $|f\rangle=|\psi_f\rangle_{12}\otimes |\bm k_3'\rangle_3\otimes |\bm k_M\rangle_M$, where $|\psi_f\rangle_{12}$ is an arbitrary eigenstate of the many-body Hamiltonian for species 1 and 2 and $|\bm k_{\rm M}\rangle_{\rm M}$ describes the state of a single molecule with momentum $\bm k_{\rm M}$.

Assuming that the binding energy $\Delta$ of the molecule is much larger than the typical single-particle energies of the initial state (chemical potential, temperature...), the momentum and energy-conservation conditions are dominated by the state of the atom 3 and of the molecule after the decay. We therefore have $\bm k_M\simeq  -\bm k'_3$ and $E_f-E_i\simeq \hbar^2 {k'_3}^2/2m+\hbar^2 k_M^2/4m-\Delta$.

The decay rate takes the form

\be
\begin{split}
\Gamma\simeq\frac{2\pi}{\hbar\Omega^3}\sum_{\psi_f,\bm k'_3}&\left|\sum_{\bm k_1,\bm k_2}\tilde g(\bm k_1-\bm k_2,\bm k'_3)_{12}\langle\psi_f|\widehat a_2(\bm k_2)\widehat a_1(\bm k_1)|0\rangle_{12}\right|^2\times\\
& \delta\left( \frac{3\hbar^2 {k'_3}^2}{4m}-\Delta\right).
\end{split}
\ee
Using the closure relation $\sum_{\psi_f}|\psi_f\rangle\langle\psi_f|=1$, we obtain
\be
\begin{split}
\Gamma\simeq\frac{2\pi}{\hbar\Omega^2}\sum_{\substack{\bm k_1,\bm k_2\\\bm k'_1,\bm k'_2}}&\chi(\bm k_1-\bm k_2,\bm k'_1-\bm k'_2)\times\\&\null_{12}\langle 0| \widehat a_1(\bm k'_1)^\dagger\widehat a_2(\bm k'_2)^\dagger\widehat a_2(\bm k_2)\widehat a_1(\bm k_1)|0\rangle_{12},
\end{split}
\ee
with
\be
\chi(\bm q,\bm q')=\frac{1}{\Omega}\sum_{\bm k'_3}\delta\left( \frac{3\hbar^2 {k'_3}^2}{4m}-\Delta\right)\tilde g(\bm q,\bm k'_3)\tilde g(\bm q',\bm k'_3)^*
\ee
Going back into real space and using the fact that thanks to momentum conservation, we must have $\bm k_1+\bm k_2=\bm k'_1+\bm k'_2$, we obtain

\be
\begin{split}
\Gamma\simeq\frac{2\pi}{\hbar}&\int d^3\bm\rho_1 d^3\bm\rho_2 G(\bm\rho_1,\bm\rho_2)\times\\ &\null_{12}\langle 0|\widehat\Psi_1^\dagger(\bm\rho_1)\widehat\Psi_2^\dagger(-\bm\rho_1)\widehat\Psi_2(\bm\rho_2)\widehat\Psi_1(-\bm\rho_2)|0\rangle_{12},
\end{split}
\ee
with
\be
G(\bm\rho,\bm\rho')=\frac{1}{\Omega^2}\sum_{\bm q,\bm q'}\chi(\bm q,\bm q')e^{-i(\bm q\cdot\bm\rho+\bm q'\cdot\bm\rho')}.
\ee
Since $G$ takes significant values for  $\rho_1,\rho_2\lesssim R^*$, we care only about the values of the correlation function $_{12}\langle 0|\widehat\Psi_1^\dagger(\bm\rho_1)\widehat\Psi_2^\dagger(\bm\rho_2)\widehat\Psi_2(\bm\rho_3)\widehat\Psi_1(\bm\rho_4)|0\rangle_{12}$ for small values of the $\rho_\alpha$.

Following  the argument proposed in \cite{werner2012generalfermions, petrov2004weakly}, the short distance behaviour is dominated by two-body physics, leading to a scaling

%In this limit it is calculated using
\be
_{12}\langle 0|\widehat\Psi_1^\dagger(\bm\rho_1)\widehat\Psi_2^\dagger(\bm\rho_2)\widehat\Psi_2(\bm\rho_3)\widehat\Psi_1(\bm\rho_4)|0\rangle_{12}\simeq \varphi(\bm\rho_1-\bm\rho_2)^*\varphi(\bm\rho_3-\bm\rho_4),
\ee
where $\varphi$ is the two-body function in a scattering state. According to Bethe-Peierls' condition, we have $\varphi(r)\simeq A/r$ at short distance, hence
\be
_{12}\langle 0|\widehat\Psi_1^\dagger(\bm\rho_1)\widehat\Psi_2^\dagger(\bm\rho_2)\widehat\Psi_2(\bm\rho_3)\widehat\Psi_1(\bm\rho_4)|0\rangle_{12}\simeq \frac{|A|^2}{|\bm\rho_1-\bm\rho_2||\bm\rho_3-\bm\rho_4|}.
\ee
The value of $|A|^2$ is obtained by taking $\bm\rho_1=\bm\rho_4$ and $\bm\rho_2=\bm\rho_4$ and, according to Tan's relation for the density-density correlation function \cite{tan2008large}, we have

\be
_{12}\langle 0|\widehat\Psi_1^\dagger(\bm\rho_1)\widehat\Psi_2^\dagger(\bm\rho_2)\widehat\Psi_2(\bm\rho_2)\widehat\Psi_1(\bm\rho_1)|0\rangle_{12}\simeq \frac{C_2}{4\pi|\bm\rho_1-\bm\rho_2|^2},
\ee
where $C_2$ is the contact. Therefore $|A|^2=C_2/4\pi$ hence

\be
\Gamma=\frac{C_2}{8\hbar }\int d^3\bm\rho_1 d^3\bm\rho_2 \frac{G(\bm\rho_1,\bm\rho_2)}{\rho_1\rho_2}.
\ee
leading to equation 1 in the main text.
\section{Feshbach resonances}

\medskip

\begin{figure}
\centerline{\includegraphics[width=\columnwidth]{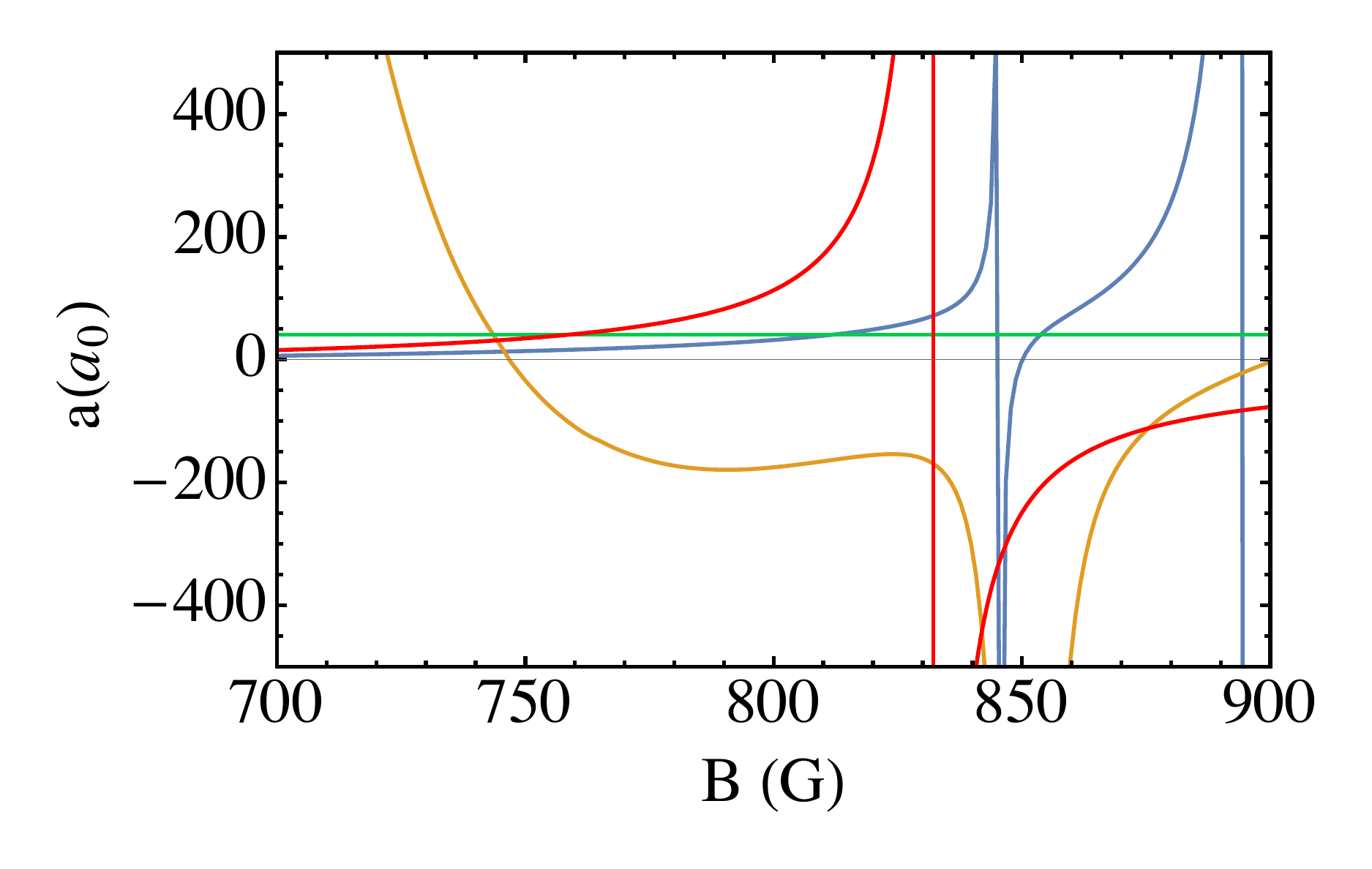}}
\caption{Magnetic fields dependence for the different scattering lengths $a_{\rm bb}$(blue), $a_{\rm ff}$(red, divided by 100), $a_{\rm bf}$(green) and effective range $r_{\rm e,b}$ (yellow) involved in our system.}
\label{FigSupp1}
\end{figure}

In Fig.~\ref{FigSupp1} we present the relevant s-wave scattering lengths characterizing the \6-\6, \7-\7,  and \6-\7 interactions ($a_{\rm ff}$, $a_{\rm bb}$, and $a_{\rm bf}$ respectively). \6 exhibits a broad Feshbach resonance at 832.18\,G \cite{Zurn:2013}.  \7-\7 interactions exhibits two Feshbach resonances located at 845.5\,G and 894\,G\,\cite{Gross:2011}. Also plotted is
the effective range for the \7-\7 interactions ($r_{e,\rm b}$) in the 700\,G-900\,G magnetic field region of interest.
 In this region $r_{\rm e,b}$ is relatively large and contributes to the \7 three-body loss dependence with magnetic field \cite{Shotan2014three}.
For the \6-\7 interaction,  the scattering length $a_{\rm bf}=40.8\,a_0$ is identical for the two \6 spin states and does not depend on the magnetic field.

\section{Loss coefficient extraction on the BEC side}

\begin{figure}
\centerline{\includegraphics[width=\columnwidth]{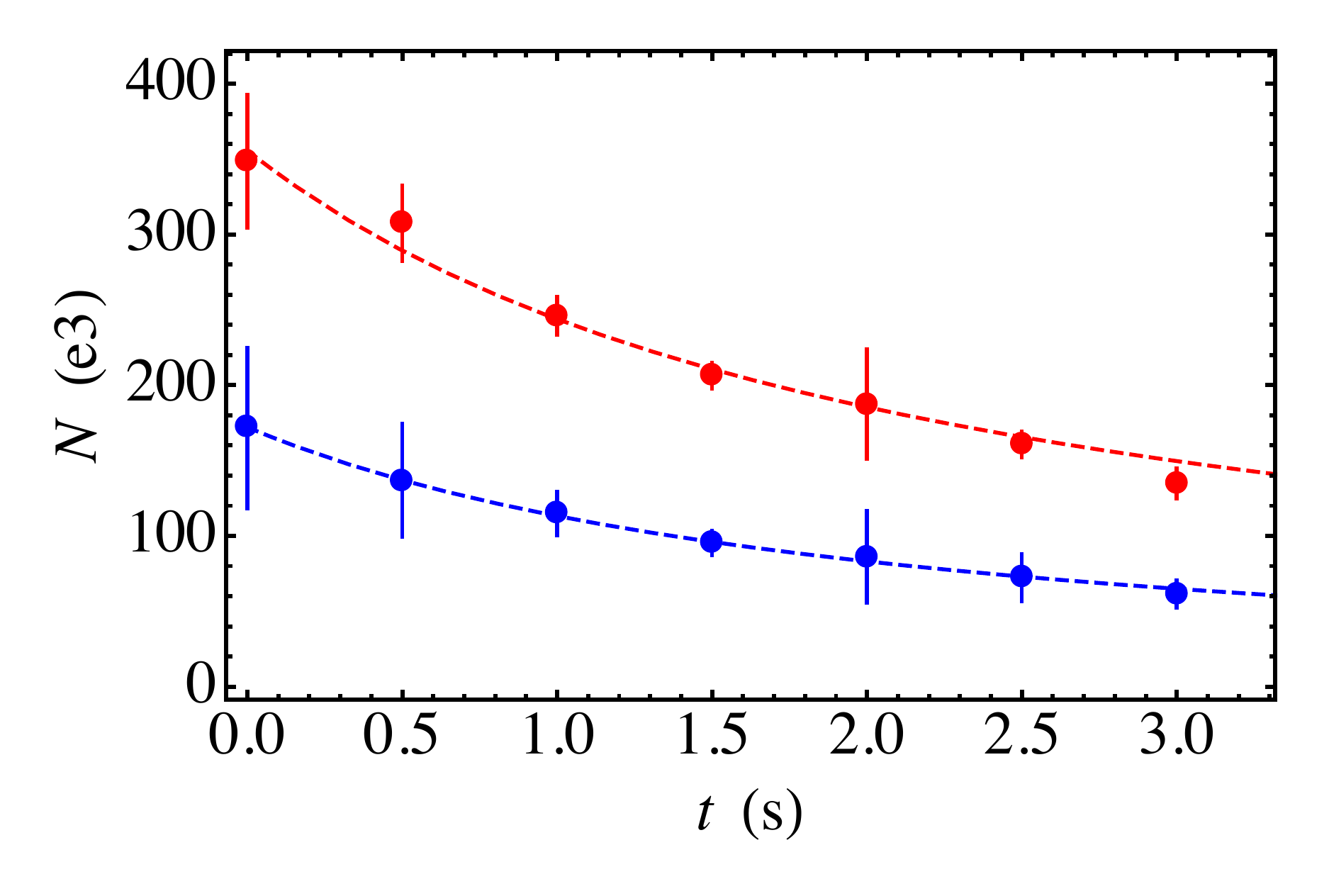}}
\caption{Example of atom losses at $B=720\,\rm G$ for a  non degenerate Bose-Fermi mixture at $T=1.25\,\mu$K. Red circles:  fermion decay. Blue circles: boson decay. Each circle is the average of  3 to 5 data points with their standard deviation. The red dashed curve is a fit of the fermion decay using Eq.\,\ref{fitfermion} to estimate $n_{\rm f}(t)$. The blue dashed curve is a fit to the boson decay using  Eq.\,\ref{fitbosonfermion} and the previously fitted $n_{\rm f}(t)$ to extract $L_{\rm bf}$.}
\label{FigSupp2}
\end{figure}

\medskip
In Fig.~\ref{FigSupp2}, we show a typical  loss rate measurement at 720\,G for a cold thermal Bose-Fermi mixture.
On the BEC side of the \6 resonance, several  processes contribute to the loss of \6 atoms: atom-dimer and dimer-dimer inelastic collisions, evaporation losses and Bose-Fermi losses, resulting in a non-trivial time dependence. On the contrary, the \7 cloud will mainly loose atoms via Bose-Fermi losses, since evaporation loss and three-body losses are negligible due to the small \7-\7 scattering length in this region of magnetic field. We thus use the boson decay rate to extract $L_{\rm bf}$  using
\be
\dot{n}_b=-L_{\rm bf}n_{\rm f} n_{\rm b}-\Gamma_v n_{\rm b}
\label{fitbosonfermion}
\ee
where the local densities follow Boltzmann distributions and $\Gamma_v=0.015\,\rm s^{-1}$ is the residual background gas loss rate measured independently. For $n_{\rm f}(t)$, we use a two-body decay function
\be
 n_{\rm f}(t)=n_{0}/(1+\alpha t)
 \label{fitfermion}
\ee

\section{Molecule fraction}
To compute the molecule fraction of the \6  cloud, $\eta=2N_{\rm m}/(N_{\rm f} +2N_{\rm m})$ we model the fermionic ensemble as a non-interacting mixture of $N_{\rm m}$ molecules and  $N_{\rm f}$ free atoms. This assumption is only valid far on the BEC side of the resonance ($n_{\rm f}a_{\rm ff}^3\ll 1$) and for a non-degenerate thermal gas. Following \cite{Chin2004thermal}, we simply write a chemical equilibrium condition  between atoms and molecules in the trap at temperature $T$:
\bea
N_{\rm f}=&2\left(\frac{k_{\rm B}T}{\hbar\bar{\omega}}\right)^3Li_3(z)\\
N_{\rm m}=&\left(\frac{k_{\rm B}T}{\hbar\bar{\omega}}\right)^3Li_3(z^2\mathbf{e}^{-E_b/k_{\rm B}T})
\eea
where $Li$ is a polylogarithm function, $z=\mathbf{e}^{\mu/k_{\rm B}T}$ the fugacity and $E_b=-\hbar^2/m_{\rm f} a_{\rm ff}^2$ the molecule's binding energy. The fugacity is calculated by imposing the total number of atoms in the trap $N_{tot}=N_{\rm f}+N_{\rm m}$.

\section{Bose-Fermi Loss coefficient at unitarity}

\begin{figure}
\centerline{\includegraphics[width=\columnwidth]{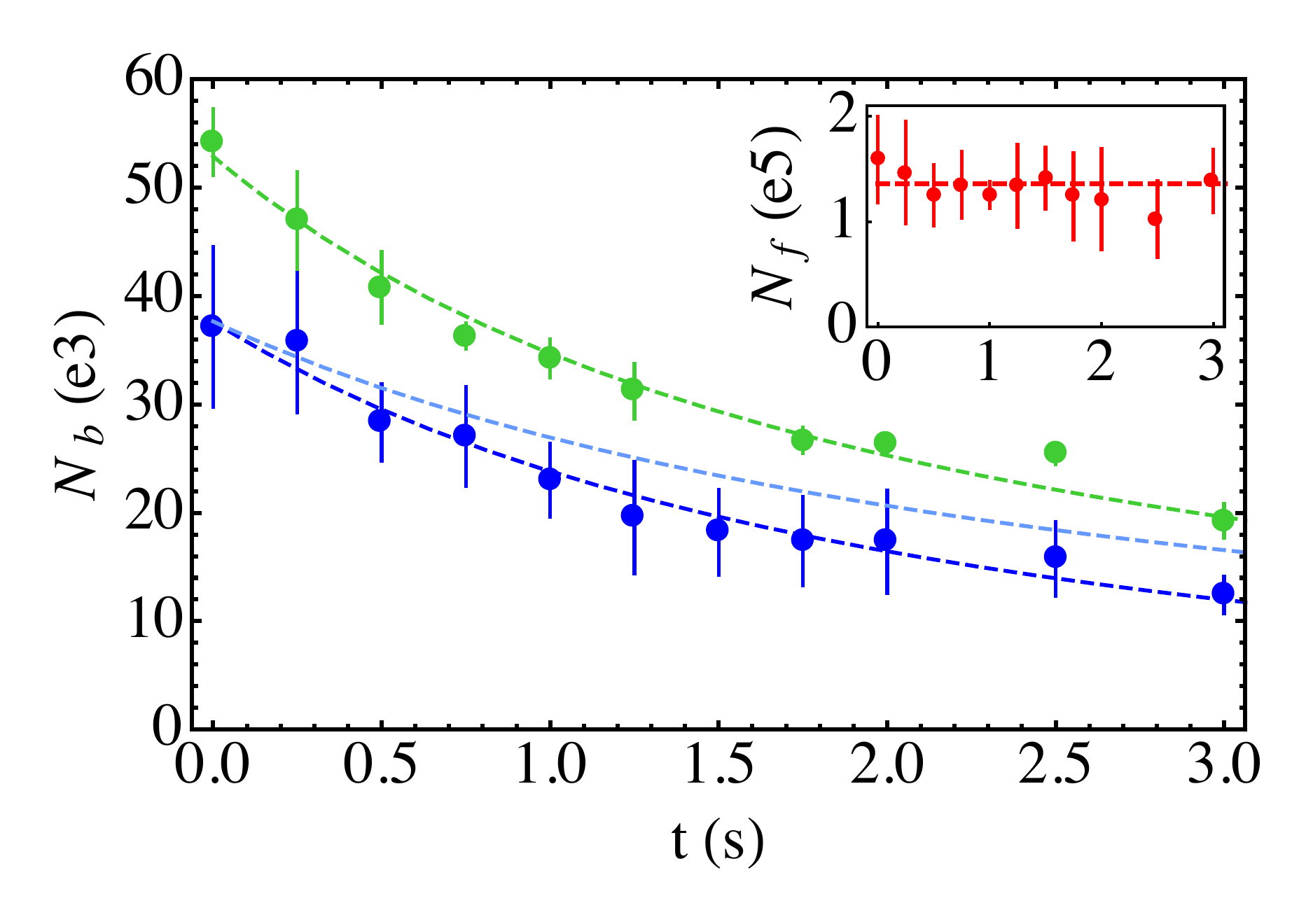}}
\caption{Example of atom loss at $B=832.1\,\rm G$ in the dual superfluid regime.  Green circles: BEC without fermions. Blue circles: BEC in presence of the fermionic superfluid. Each circle is the average of 3 to 5 data points with their standard deviation. Green dashed curve: fit to the decay of the BEC alone using Eq.~\ref{B3body} providing the three body loss coefficient $L_{\rm b}$. Blue dashed curve: fit to the BEC with fermions using Eq.~\ref{Gammabf} which gives $\Gamma_{\rm bf}=0.14(4)\,s^{-1}$. Light blue dashed curve: expected BEC decay without Bose-Fermi losses. Inset: the number of \6 atoms for the same time duration (red circles). As it is nearly constant we use the mean number of \6 atoms shown as a red dashed line to compute the peak density of the fermionic superfluid during the losses.}
\label{FigSupp3}
\end{figure}

In Fig.~\ref{FigSupp3}, we show a typical loss rate measurement in the dual superfluid regime at unitarity. As stated in the main text the three-body recombination in the Bose gas itself also contribute to the boson decay.
In order to measure the Bose-Fermi loss rate $\Gamma_{\rm bf}$ we first measure the three-body loss coefficient $L_{\rm b}$ for a BEC alone using the following equation,
\be
\dot{n}_b=-L_{\rm b}n_{\rm b}^3-\Gamma_v n_{\rm b}
\label{B3body}
\ee
We restrict the measurement over a period of time for which the thermal fraction surrounding the BEC is not visible. We thus assume that the \7 cloud density is given by a Thomas-Fermi distribution.
Knowing $L_{\rm b}$, we extract $\Gamma_{\rm bf}$ for a BEC in presence of fermions in the same trap using
\be
\dot{n}_b=-L_{\rm b}n_{\rm b}^3-\Gamma_v n_{\rm b}-\Gamma_{\rm bf}n_{\rm b}.
\label{Gammabf}
\ee
Since $n_{\rm f}$ does not vary significantly over the  measurement duration, see Fig.\ref{FigSupp3}, $\Gamma_{\rm bf}$ is now assumed constant.
\medskip

\section{Reduction of the Bose-Fermi losses due to the finite size of the BEC}

In the dual superfluid regime, the bosonic sample is not a perfect point-like impurity probing the central density of the Fermi gas.
The finite size of the BEC leads to a slight reduction of the Bose-Fermi losses that we compute using the local density approximation:
\be
\frac{\langle n_{\rm f}^{4/3}(\mathbf{r})\rangle_{\rm BEC}}{n_{\rm f}^{4/3}(0)}=\frac{\int d^3r n_{\rm b}(\mathbf{r})n_{\rm f}^{4/3}(\mathbf{r})}{n_{\rm f}^{4/3}(0)\int d^3r n_{\rm b}(\mathbf{r})}
\ee
where $n_{\rm b}$ and $n_{\rm f}$ are respectively the boson and fermion densities. Introducing the Thomas-Fermi radii $R_{\rm TF, b}$ and $R_{\rm TF, f}$, we find
\be
\frac{\langle n_{\rm f}^{4/3}(\mathbf{r})\rangle_{\rm BEC}}{n_{\rm f}^{4/3}(0)}=1-\frac{6}{7}\left(\frac{R_{\rm TF,b}}{R_{\rm TF, f}}\right)^2+\frac{5}{21}\left(\frac{R_{\rm TF, b}}{R_{\rm TF, f}}\right)^4.
\ee
With $R_{\rm TF, b}=0.4 R_{\rm TF, f}$, the reduction factor amounts to 0.9.


\begin{thebibliography}{0}
\expandafter\ifx\csname natexlab\endcsname\relax\def\natexlab#1{#1}\fi
\expandafter\ifx\csname bibnamefont\endcsname\relax
  \def\bibnamefont#1{#1}\fi
\expandafter\ifx\csname bibfnamefont\endcsname\relax
  \def\bibfnamefont#1{#1}\fi
\expandafter\ifx\csname citenamefont\endcsname\relax
  \def\citenamefont#1{#1}\fi
\expandafter\ifx\csname url\endcsname\relax
  \def\url#1{\texttt{#1}}\fi
\expandafter\ifx\csname urlprefix\endcsname\relax\def\urlprefix{URL }\fi
\providecommand{\bibinfo}[2]{#2}
\providecommand{\eprint}[2][]{\url{#2}}

\end{thebibliography}


\begin{thebibliography}{10}

\bibitem{bloch2008many}
I.~Bloch, J.~Dalibard, and W.~Zwerger.
\newblock {Many-body physics with ultracold gases}.
\newblock {\em Rev. Mod. Phys.}, 80(3):885--964, 2008.

\bibitem{zwerger2012BCSBEC}
W.~Zwerger, editor.
\newblock {\em The BCS-BEC Crossover and the Unitary Fermi Gas}, volume 836 of
  {\em Lecture Notes in Physics}.
\newblock Springer, Berlin, 2012.

\bibitem{Inguscio2016quantum}
M.~Inguscio, W.~Ketterle, S.~Stringari, and G.~Roati, editors.
\newblock {\em Quantum Matter at Ultralow Temperatures}, volume 191 of {\em
  Proceedings of the International School of Physics "Enrico Fermi"}.
\newblock IOS Press, Varenna, 2016.

\bibitem{choi2016exploring}
Jae-yoon Choi, Sebastian Hild, Johannes Zeiher, Peter Schau{\ss}, Antonio
  Rubio-Abadal, Tarik Yefsah, Vedika Khemani, David~A Huse, Immanuel Bloch, and
  Christian Gross.
\newblock Exploring the many-body localization transition in two dimensions.
\newblock {\em Science}, 352:1547, 2016.

\bibitem{makotyn2014universal}
Philip Makotyn, Catherine~E Klauss, David~L Goldberger, EA~Cornell, and
  Deborah~S Jin.
\newblock Universal dynamics of a degenerate unitary {B}ose gas.
\newblock {\em Nature Physics}, 10(2):116--119, 2014.

\bibitem{chevy2016strongly}
F.~Chevy and C.~Salomon.
\newblock Strongly correlated {B}ose gases.
\newblock {\em Journal of Physics B: Atomic, Molecular and Optical Physics},
  49(19):192001, 2016.

\bibitem{syassen2008strong}
Niels Syassen, Dominik~M Bauer, Matthias Lettner, Thomas Volz, Daniel Dietze,
  Juan~J Garcia-Ripoll, J~Ignacio Cirac, Gerhard Rempe, and Stephan D{\"u}rr.
\newblock Strong dissipation inhibits losses and induces correlations in cold
  molecular gases.
\newblock {\em Science}, 320(5881):1329--1331, 2008.

\bibitem{daley2009atomic}
A.J. Daley, J.M. Taylor, S.~Diehl, M.~Baranov, and P.~Zoller.
\newblock Atomic three-body loss as a dynamical three-body interaction.
\newblock {\em Phys. Rev. Lett.}, 102(4):040402, 2009.

\bibitem{zhu2014suppressing}
Bihui Zhu, Bryce Gadway, Michael Foss-Feig, Johannes Schachenmayer, ML~Wall,
  Kaden~RA Hazzard, Bo~Yan, Steven~A Moses, Jacob~P Covey, Deborah~S Jin,
  et~al.
\newblock Suppressing the loss of ultracold molecules via the continuous
  quantum zeno effect.
\newblock {\em Phys. Rev. Lett.}, 112(7):070404, 2014.

\bibitem{kraemer2006evidence}
T.~Kraemer, M.~Mark, P.~Waldburger, JG~Danzl, C.~Chin, B.~Engeser, AD~Lange,
  K.~Pilch, A.~Jaakkola, H.C. N{\"a}gerl, et~al.
\newblock {Evidence for Efimov quantum states in an ultracold gas of caesium
  atoms}.
\newblock {\em Nature}, 440(7082):315--318, 2006.

\bibitem{Tan2008energetics}
Shina Tan.
\newblock Energetics of a strongly correlated fermi gas.
\newblock {\em Annals of Physics}, 323(12):2952 -- 2970, 2008.

\bibitem{tan2008large}
S.~Tan.
\newblock {Large momentum part of a strongly correlated {F}ermi gas}.
\newblock {\em Ann. Phys.}, 323(12):2971--2986, 2008.

\bibitem{olshanii2003short}
Maxim Olshanii and Vanja Dunjko.
\newblock Short-distance correlation properties of the lieb-liniger system and
  momentum distributions of trapped one-dimensional atomic gases.
\newblock {\em Phys. Rev. Lett.}, 91(9):090401, 2003.

\bibitem{d2008suppression}
J.P. D’Incao and B.D. Esry.
\newblock Suppression of molecular decay in ultracold gases without {F}ermi
  statistics.
\newblock {\em Phys. Rev. Lett.}, 100(16):163201, 2008.

\bibitem{zirbel2008collisional}
J.J. Zirbel, K.-K. Ni, S.~Ospelkaus, J.P. D’Incao, C.E. Wieman, J.~Ye, and D.S.
  Jin.
\newblock Collisional stability of fermionic {F}eshbach molecules.
\newblock {\em Phys. Rev. Lett.}, 100(14):143201, 2008.

\bibitem{dincao2009short}
J~P D'Incao, Chris~H Greene, and B~D Esry.
\newblock The short-range three-body phase and other issues impacting the
  observation of efimov physics in ultracold quantum gases.
\newblock {\em Journal of Physics B: Atomic, Molecular and Optical Physics},
  42(4):044016, 2009.

\bibitem{spiegelhalder2009collisional}
F.~M. Spiegelhalder, A.~Trenkwalder, D.~Naik, G.~Hendl, F.~Schreck, and
  R.~Grimm.
\newblock {Collisional Stability of $^{40}${K} Immersed in a Strongly
  Interacting {F}ermi Gas of $^6${L}i}.
\newblock {\em Phys. Rev. Lett.}, 103(22):223203, Nov 2009.

\bibitem{Khramov2012dynamics}
Alexander~Y. Khramov, Anders~H. Hansen, Alan~O. Jamison, William~H. Dowd, and
  Subhadeep Gupta.
\newblock Dynamics of feshbach molecules in an ultracold three-component
  mixture.
\newblock {\em Phys. Rev. A}, 86:032705, Sep 2012.

\bibitem{Kagan1985effect}
Y.~{Kagan}, B.~V. {Svistunov}, and G.~V. {Shlyapnikov}.
\newblock {Effect of Bose condensation on inelastic processes in gases}.
\newblock {\em Soviet Journal of Experimental and Theoretical Physics Letters},
  42:209, August 1985.

\bibitem{petrov2004weakly}
D.S. Petrov, C.~Salomon, and G.V. Shlyapnikov.
\newblock {Weakly bound dimers of fermionic atoms}.
\newblock {\em Phys. Rev. Lett.}, 93(9):090404, 2004.

\bibitem{astrakharchik2004eq}
G.E. Astrakharchik, J.~Boronat, J.~Casulleras, and S.~Giorgini.
\newblock {Equation of state of a {F}ermi gas in the BEC-BCS crossover: A
  quantum Monte Carlo study}.
\newblock {\em Phys. Rev. Lett.}, 93(20):200404, 2004.

\bibitem{navon2010equation}
Nir Navon, Sylvain Nascimb{\`e}ne, Fr{\'e}d{\'e}ric Chevy, and Christophe
  Salomon.
\newblock The equation of state of a low-temperature fermi gas with tunable
  interactions.
\newblock {\em Science}, 328(5979):729--732, 2010.

\bibitem{sagi2012measurement}
Yoav Sagi, Tara~E Drake, Rabin Paudel, and Deborah~S Jin.
\newblock Measurement of the homogeneous contact of a unitary fermi gas.
\newblock {\em Physical review letters}, 109(22):220402, 2012.

\bibitem{kuhnle2011temperature}
E.D. Kuhnle, S.~Hoinka, P.~Dyke, H.~Hu, P.~Hannaford, and C.J. Vale.
\newblock Temperature dependence of the universal contact parameter in a
  unitary {F}ermi gas.
\newblock {\em Phys. Rev. Lett.}, 106(17):170402, 2011.

\bibitem{hoinka2013}
Sascha Hoinka, Marcus Lingham, Kristian Fenech, Hui Hu, Chris~J. Vale,
  Joaqu\'{\i}n~E. Drut, and Stefano Gandolfi.
\newblock Precise determination of the structure factor and contact in a
  unitary fermi gas.
\newblock {\em Phys. Rev. Lett.}, 110:055305, Jan 2013.

\bibitem{SuppMat}
See supplementary information.

\bibitem{Ferrier2014Mixture}
I~Ferrier-Barbut, M.~Delehaye, S.~Laurent, A.T. Grier, M.~Pierce, B.S Rem,
  F.~Chevy, and C.~Salomon.
\newblock {A mixture of {B}ose and Fermi superfluids}.
\newblock {\em Science}, 345:1035--1038, 2014.

\bibitem{Zurn:2013}
G.~Z{\"u}rn, T.~Lompe, A.~N. Wenz, S.~Jochim, P.~S. Julienne, and J.~M. Hutson.
\newblock {Precise Characterization of $^{6}$Li Feshbach Resonances Using
  Trap-Sideband-Resolved RF Spectroscopy of Weakly Bound Molecules}.
\newblock {\em Phys. Rev. Lett.}, 110(13):135301, 2013.

\bibitem{Chin2004thermal}
Cheng Chin and Rudolf Grimm.
\newblock Thermal equilibrium and efficient evaporation of an ultracold
  atom-molecule mixture.
\newblock {\em Phys. Rev. A}, 69:033612, Mar 2004.

\bibitem{fedichev1996three}
P.~O. Fedichev, M.~W. Reynolds, and G.~V. Shlyapnikov.
\newblock Three-body recombination of ultracold atoms to a weakly bound
  $\mathit{s}$ level.
\newblock {\em Phys. Rev. Lett.}, 77:2921--2924, Sep 1996.

\bibitem{Shotan2014three}
Zav Shotan, Olga Machtey, Servaas Kokkelmans, and Lev Khaykovich.
\newblock Three-body recombination at vanishing scattering lengths in an
  ultracold bose gas.
\newblock {\em Phys. Rev. Lett.}, 113:053202, Jul 2014.

\bibitem{ku2012revealing}
Mark~JH Ku, Ariel~T Sommer, Lawrence~W Cheuk, and Martin~W Zwierlein.
\newblock Revealing the superfluid lambda transition in the universal
  thermodynamics of a unitary fermi gas.
\newblock {\em Science}, 335(6068):563--567, 2012.

\bibitem{roy2016two}
Richard Roy, Alaina Green, Ryan Bowler, and Subhadeep Gupta.
\newblock Two-element mixture of bose and fermi superfluids.
\newblock {\em arXiv preprint arXiv:1607.03221}, 2016.

\bibitem{yao2016observation}
Xing-Can Yao, Hao-Ze Chen, Yu-Ping Wu, Xiang-Pei Liu, Xiao-Qiong Wang, Xiao
  Jiang, Youjin Deng, Yu-Ao Chen, and Jian-Wei Pan.
\newblock Observation of two-species vortex lattices in a mixture of
  mass-imbalance bose and fermi superfluids.
\newblock {\em arXiv preprint arXiv:1606.01717}, 2016.

\bibitem{Ikemachi2016all-optical}
T.~Ikemachi, A.~Ito, .Y~Aratake, Y.~Chen, M.~Koashi, M.~Kuwata-Gonomaki, and
  M.~Horikoshi.
\newblock All-optical production of a superfluid bose-fermi mixture of
  $^6\mathrm{Li}$ and $^7\mathrm{Li}$.
\newblock {\em arXiv preprint arXiv:1606.09404}, 2016.

\bibitem{Braaten2009three}
Eric Braaten, H.-W. Hammer, Daekyoung Kang, and Lucas Platter.
\newblock Three-body recombination of $^{6}\mathrm{Li}$ atoms with large
  negative scattering lengths.
\newblock {\em Phys. Rev. Lett.}, 103:073202, Aug 2009.

\bibitem{Ottenstein2008collisional}
T.~B. Ottenstein, T.~Lompe, M.~Kohnen, A.~N. Wenz, and S.~Jochim.
\newblock Collisional stability of a three-component degenerate fermi gas.
\newblock {\em Phys. Rev. Lett.}, 101:203202, Nov 2008.

\bibitem{Sagi2012}
Yoav Sagi, Tara~E. Drake, Rabin Paudel, and Deborah~S. Jin.
\newblock Measurement of the homogeneous contact of a unitary fermi gas.
\newblock {\em Phys. Rev. Lett.}, 109:220402, Nov 2012.

\bibitem{Fletcher2016}
R.~J. {Fletcher}, R.~{Lopes}, J.~{Man}, N.~{Navon}, R.~P. {Smith}, M.~W.
  {Zwierlein}, and Z.~{Hadzibabic}.
\newblock {Two and Three-body Contacts in the Unitary Bose Gas}.
\newblock {\em ArXiv e-prints}, 2016.

\bibitem{Du2009inelastic}
X.~Du, Y.~Zhang, and J.~E. Thomas.
\newblock Inelastic collisions of a fermi gas in the bec-bcs crossover.
\newblock {\em Phys. Rev. Lett.}, 102:250402, Jun 2009.

\end{thebibliography}

\begin{thebibliography}{1}

\bibitem{werner2012generalfermions}
F{\'e}lix Werner and Yvan Castin.
\newblock General relations for quantum gases in two and three dimensions:
  Two-component fermions.
\newblock {\em Phys. Rev. A}, 86(1):013626, 2012.

\bibitem{petrov2004weakly}
D.S. Petrov, C.~Salomon, and G.V. Shlyapnikov.
\newblock {Weakly bound dimers of fermionic atoms}.
\newblock {\em Phys. Rev. Lett.}, 93(9):090404, 2004.

\bibitem{tan2008large}
S.~Tan.
\newblock {Large momentum part of a strongly correlated {F}ermi gas}.
\newblock {\em Ann. Phys.}, 323(12):2971--2986, 2008.

\bibitem{Zurn:2013}
G.~Z{\"u}rn, T.~Lompe, A.~N. Wenz, S.~Jochim, P.~S. Julienne, and J.~M. Hutson.
\newblock {Precise Characterization of $^{6}$Li Feshbach Resonances Using
  Trap-Sideband-Resolved RF Spectroscopy of Weakly Bound Molecules}.
\newblock {\em Phys. Rev. Lett.}, 110(13):135301, 2013.

\bibitem{Gross:2011}
N.~Gross, Z.~Shotan, O.~Machtey, S.J.J.M.F Kokkelmans, and L.~Khaykovich.
\newblock {Study of Efimov Physics in two nuclear-spin sublevels of 7 Li}.
\newblock {\em Comptes Rendus Physique}, 12(1):4--12, 2011.

\bibitem{Shotan2014three}
Zav Shotan, Olga Machtey, Servaas Kokkelmans, and Lev Khaykovich.
\newblock Three-body recombination at vanishing scattering lengths in an
  ultracold bose gas.
\newblock {\em Phys. Rev. Lett.}, 113:053202, Jul 2014.

\bibitem{Chin2004thermal}
Cheng Chin and Rudolf Grimm.
\newblock Thermal equilibrium and efficient evaporation of an ultracold
  atom-molecule mixture.
\newblock {\em Phys. Rev. A}, 69:033612, Mar 2004.

\end{thebibliography}
\end{document}